# Electron Transport across Vertical Silicon / MoS$_2$ / Graphene Heterostructures: Towards Efficient Emitter Diodes for Graphene-Base Hot Electron Transistors

Melkamu Belete[1], Olof Engström[2], Sam Vaziri[3], Gunther Lippert[4], Mindaugas Lukosius[4], Satender Kataria[1,*], Max C. Lemme[1,2,*]

[1]RWTH Aachen University, Faculty of Electrical Engineering and Information Technology, Chair of Electronic Devices, Otto-Blumenthal-Str. 2, 52074 Aachen, Germany

[2]AMO GmbH, Advanced Microelectronic Center Aachen, Otto-Blumenthal-Str. 25, 52074 Aachen, Germany

[3]Stanford University, Department of Electrical Engineering, Stanford, CA 94305, USA

[4]IHP - Leibniz-Institut für innovative Mikroelektronik, IHP GmbH, Frankfurt (Oder), Germany

Email: satender.kataria@eld.rwth-aachen.de, lemme@amo.de

**Abstract:**

Heterostructures comprising of silicon, molybdenum disulfide (MoS$_2$) and graphene are investigated with respect to the vertical current conduction mechanism. The measured current-voltage (I-V) characteristics exhibit temperature dependent asymmetric current, indicating thermally activated charge carrier transport. The data is compared and fitted to a current transport model that confirms thermionic emission as the responsible transport mechanism across the devices. Theoretical calculations in combination with the experimental data suggest that the heterojunction barrier from Si to MoS$_2$ is linearly temperature dependent for T = 200 to 300 K with a positive temperature coefficient. The temperature dependence may be attributed to a change in band gap difference between Si and MoS$_2$, strain at the Si/MoS$_2$ interface or different electron effective masses in Si and MoS$_2$, leading to a possible entropy change stemming from variation in density of states as electrons move from Si to MoS$_2$. The low barrier formed between Si and MoS$_2$ and the resultant thermionic emission demonstrated here makes the present devices potential candidates as the emitter diode of graphene-base hot electron transistors for future high-speed electronics.

## I. Introduction

Hot electron transistors (HETs) have been proposed first by Mead in the 1960's as potential high performance electron devices [1]. Such transistors rely on vertical device structures with cross-plane transport of high energy electrons (hot electrons). The first HETs comprised of metal emitter, base, and collector, isolated from each other by thin oxide layers. The cutoff frequency in these devices is limited by the base transit time. While thinning down the metallic base mitigates this issue, it increases the base resistance resulting in high RC delay and self-bias crowding. Graphene-base transistors (GBTs), where graphene replaces the metallic base electrodes, have been proposed to exploit the high conductivity and ultra-thinness of graphene as the base material in HETs to minimize the base transit time and achieve high cutoff frequencies [2–10]. Although experimental demonstration of GBTs is limited to DC characteristics, simulations clearly show that the performance greatly depends on the properties of the injection barrier that isolates the emitter and the base. In fact, high level on-state collector currents ($I_{ON}$) can be achieved only by choosing injection barriers that form relatively small conduction band (CB) offsets with respect to the emitter [11–13]. Thus, vertical heterostructures with low barriers, similar to compound semiconductor structures investigated by Heiblum et al. [14,15], have been proposed to enable high frequency performance reaching (theoretically) the THz regime [11–13]. Illustration of the structure and operation of a GBT with $MoS_2$ as the emission barrier is shown by the schematics in Fig. S1.

The two-dimensional (2D) material molybdenum disulfide ($MoS_2$) is one of the most explored 2D transition metal dichalcogenides (TMDs) due to its electronic properties that have made it a potential choice for future nanoelectronic, optoelectronic and flexible electronic devices [5,16–22]. Unlike graphene, $MoS_2$ is known for its intrinsic semi-conducting behavior and sizeable band-gap (i.e. 1.3 eV in bulk and as high as 2.16 eV in mono-layer) [20,23]. According to theory, MoS$_2$ can provide a desirable low barrier for carrier injection in GBTs in combination with highly doped silicon emitters and graphene base electrodes. The electron affinity of bulk MoS$_2$ is reported to be 4 eV [24,25] indicating the small band offset it would make with respect to Si that has 4.05 eV. Also, based on literature values of the band gaps of Si (1.1 eV as mentioned elsewhere) and bulk MoS$_2$ (1.3 eV [20]), there would only be ~ 0.2 eV to be distributed between the electron and hole barriers at the Si-MoS$_2$ interface. According to our previous study, the electron barrier is much smaller than the hole barrier [26]. Hence, the CB offset at the Si/MoS$_2$ interface should be small enough to make MoS$_2$ an efficient injection barrier in devices like GBTs. Charge transport across vertical "metal/exfoliated-MoS$_2$/metal" structures has been reported to involve Fowler-Nordheim and thermal injections at high and low electric fields, respectively [27]. However, detailed studies on the vertical transport properties across Si/MoS$_2$ interfaces are not available. In this work, we experimentally investigate vertical electron transport across vapor-phase grown layered MoS$_2$ sandwiched between highly doped silicon and graphene. The Si/MoS$_2$/graphene heterostructures were fabricated with a scalable process and characterized electrically. As GBTs operate in the forward bias regime, where hot electrons from the n$^+$-Si emitter are injected into the graphene base and then to the collector, the focus of this work is on the electron transport in the forward-biased regime. The charge carrier transport mechanism is explained through calculations and analyses based on the measured data. Furthermore, electron barrier height values were determined using two different methods: (1) From thermal activation plots and (2) by fitting voltage dependent barrier heights to the current-voltage (I-V) characteristics measured at different temperatures.

## II.  Experimental Section

## A. Device fabrication

The semiconductor-semiconductor-graphene (SSG) heterostructures were fabricated as follows: first, Si active areas of various sizes were defined on Si (100) wafers through photolithography followed by Si etching to form trenches. The trenches were then filled with thick silicon oxide ($SiO_2$) layer deposited by high density plasma chemical vapor deposition (HDPCVD) technique. The $SiO_2$ layer serves as a shallow trench isolation (STI) that separates neighboring devices and also helps to avoid direct leakage paths from the metal pads to the underlying Si. After $SiO_2$ deposition, chemical mechanical polishing (CMP) was employed to planarize the wafer surface. Then, to create locally doped $n^+$ Si active regions, phosphorous ion implantation was carried out using silicon nitride as a hard mask. After dicing the wafers into small chips (1.5 cm x 1.5 cm) and performing a standard cleaning procedure, photolithography was used to define larger windows covering the Si active areas. Then, native oxide was removed from the active areas using (7:1) buffered oxide etch (BOE) solution, followed by deposition of ~ 5 nm molybdenum (Mo) films using e-beam evaporation. After a lift-off process to complete the Mo film patterning, a thermally assisted conversion (TAC) of the Mo films was carried out inside a chemical vapor deposition (CVD) furnace to achieve ~ 15 nm $MoS_2$ films as confirmed by atomic force microscopy (AFM) inspection (Fig. S2 in supplementary information). Also, transmission electron microscopy (TEM) investigation was carried out to inspect the structural formation of the $MoS_2$ film. The resulting TEM cross-section image and the corresponding electron diffraction pattern indicate the polycrystalline nature of the film (Fig. S3 in supplementary information). Next, a CVD grown single layer graphene (SLG) was transferred onto the Si/$MoS_2$ target samples using a polymer-assisted wet chemical etching transfer technique [28,29]. Afterwards, the SLG was patterned using a step of photolithography followed by reactive ion etching (RIE) in oxygen plasma. Then, metal contacts were formed at both ends of the SLG through a sequence of photolithography, evaporation of (20/120) nm Cr/Au stack and lift-off processes. Finally, the device fabrication was concluded by depositing Cr/Au metal back-contacts to the Si substrates. Schematics in isometric view of the SSG device are presented in Figs. 1a and 1b. In addition, top view optical micrograph of the as-fabricated device is shown in Fig. 1c. To inspect the presence and quality of the $MoS_2$ and SLG layers, Raman spectroscopy was performed on the samples and the corresponding Raman spectra are presented in Fig. 1d, confirming a 2H-$MoS_2$ phase formation and the monolayer nature of the graphene.

**B. Modeling thermionic emission across small electron barriers**

The early understanding of thermionic emission of charge carriers from solid state materials began in the 1930s during the development of vacuum tube electronics. Later in 1949, Herring and Nichols summarized the basic formulation of a theory describing the flow of electrons from metals into vacuum [30]. With some refinements over the years, the theory is still being used to describe thermionic emission between metals and semiconductors [31–36]. In the former case, the electron needs to overcome a barrier constituted by the metal work function which allows to assume isotropic electronic properties, parabolic energy bands and thus Maxwellian velocity distributions of electrons on both sides of the barrier. For a metal/silicon Schottky structure, a similar assumption can be adopted for the metal while ellipsoidal constant energy surfaces need to be considered for the silicon CB [37]. This is required in order to maintain the Maxwellian velocity distribution of charge carriers as a basis for the description of thermionic emission.

In this paper, we investigate electron transport from a highly doped Si into 15 nm layered $MoS_2$ by assuming a Maxwellian distribution of electrons in the Si passing over a voltage- and temperature-dependent barrier, $\Phi_B$ that varies from ~ 0.3 eV down to 0 eV. Schematic diagrams illustrating the charge distribution in the present system and the associated band alignments are shown in Fig. 2. If the right hand side of the Si/MoS$_2$ barrier geometry in Fig. 2 were vacuum, the electrons from Si would continue into a new isotropic and parabolic energy band. However, in the present sample, they enter an indirect gap polycrystalline MoS$_2$ with more than one possible CB.[35,36,38] During the charge carrier transfer process, electrons with velocity components in the *x*-direction and energy larger than $\Phi_B$ are injected into the MoS$_2$ layer. At high voltages, $\Phi_B$ approaches values lower than the 3kT limit below which the Boltzmann statistics is not valid. Therefore, Fermi statistics is needed to be used in our calculations to obtain more reliable current data. The injected current is considered proportional to the product *f(E)g(E)* of the Fermi-function, *f(E)*, and the density of states, *g(E)*, integrated along energy, E, of the Si CB from the barrier energy value up to infinity. Hence, the current density, *J*, becomes

$$J = BT \int_{\Phi_B}^{\infty} \left[ \frac{E^{1/2}}{1+exp\left(\frac{E-E_F}{kT}\right)} \right] dE, \qquad (1)$$

Where *B* is a constant, $E_F$ is the Fermi-level of Si, *k* is the Boltzmann's constant and *T* is temperature. The solution of the integral in Eq. 1, gives rise to an additional pre-factor *T*, such that *J* becomes proportional to $T^2$. The assumption, which Eq. 1 is based on, requires that the electron transmission probability is independent of electron energy and that the two materials have similar density of states as a function of energy. The Fermi-level of the Si emitter is close to the CB edge due to the high doping level and that makes it possible to treat the structure similar to a Schottky diode. Therefore, for $\Phi_B$ larger than about 3*kT*, the standard Richardson expression given by [33,34]

$$J = A^* T^2 \exp\left(\frac{-\Phi_B}{kT}\right), \qquad (2)$$

where *A\** is a pre-exponential coefficient known as the effective Richardson constant, applies.

As will be demonstrated later in the discussion section, the heterojunction barrier in the present case has both temperature and bias dependence that can be described as

$$\Phi_B(V,T) = \Phi_{B0}(V) + \alpha(V,T)T, \qquad (3)$$

Where $\Phi_{B0}(V)$ is a bias-dependent barrier height and $\alpha(V)$ is a proportionality factor at *T* = 0 K provided that *α* is temperature independent. Combining the Richardson's expression in Eq.(2) with Eq. (3), one can get

$$\frac{J}{T^2} = A^* exp\left[-\left(\frac{\alpha(V)}{k} + \frac{\Phi_{B0}(V)}{kT}\right)\right]. \qquad (4)$$

Hence, an Arrhenius plot of $\ln\left(\frac{J}{T^2}\right)$ vs $\frac{1}{kT}$ would give a slope determined by $\Phi_{B0}$.

### C. Influence of charge on the MoS$_2$ band diagram

In the present structure, a small electron barrier (i.e. < 100 meV) forms at the Si-MoS$_2$ interface at thermal equilibrium. Hence, electrons injected from the Si into MoS$_2$ face this low energy barrier (EB1) provided that the MoS$_2$ layer is neutral (Fig. 2a). However, as was found in our earlier work [26], negative charge exists inside the MoS$_2$ bulk that influences the shape of the MoS$_2$ CB by inducing an additional barrier (EB2) for electrons (Fig. 2b). This moves the barrier maximum from the interface to a point on the CB edge on the MoS$_2$ layer. The influence of the aforementioned charge on the electron barrier can be modeled using the Poisson equation. To maintain charge conservation, the net charge in the Si/MoS$_2$/Graphene(Gr) structure should be zero. This implies that the MoS$_2$ bulk charge, $Q_B(x)$, is compensated by charges at the Si/MoS$_2$ and MoS$_2$/G interfaces as illustrated in the schematic diagram in Fig. 2c. If the charge at the Si-MoS2 interface, $Q_{Si}$, is a fraction, $a$, of the total bulk charge, $Q_B$, so that

$$Q_{Si} = a\, Q_B(x), \qquad (5)$$

then the charge at the MoS2/Gr interface would be

$$Q_{Gr} = (1 - a) Q_B(x), \qquad (6)$$

where $x$ is the distance from the Si interface into the MoS2 bulk. For an arbitrary depth distribution, $N(x)$, of negative ions inside MoS2, the total bulk charge would be

$$Q_B(x) = -q \int_0^{x_d} N(x) dx, \qquad (7)$$

where the "-" sign is due to the negative bulk charge in the present case. For a known depth distribution of the bulk charge, Eqs. (5 – 7) can be used to derive an expression for the electric field, *F(x)*, as

$$F(x) = \frac{1}{\varepsilon \varepsilon_0} (Q_B(x) + Q_{Si}) + \frac{(\Phi_{Gr} - \Phi_{Si})}{x_d}, \qquad (8)$$

where $\Phi_G$ and $\Phi_{Si}$ are the work functions of graphene and Si, respectively. Once the electric field is obtained using Eq. (8), the bending of the MoS2 CB, $E_C$[eV], can be calculated using the expression

$$E_C(x) = \int_0^{x_d} F(x) dx. \qquad (9)$$

Accounting for the effect of image force barrier lowering due to image charges building up in the Si as electrons leave, Eq. 9 can be rewritten as

$$E_C(x) = \left[\int_0^{x_d} F(x) dx\right] - \frac{q}{16\pi\varepsilon\varepsilon_0}\left(\frac{1}{x}\right) \qquad (10)$$

where $\varepsilon_0$ is the permitivity of vacuum and $\varepsilon$ is the electronic dielectric constant of MoS2. For the $E_C(x)$ calculations, an electronic dielectric constant value of $\varepsilon_i$ = 3 was used based on the following considerations. High bias voltages result in lowering of the injection barrier maximum (EB2). This increases the number of lower thermal energy electrons that are injected into the MoS2 and this in turn lowers the average electron velocity in the MoS2 crystal. As the dielectric constant depends on carrier velocity, it is therefore possible that the electronic dielectric constant of the MoS$_2$ in the present case approaches the static value [26].

## III. Results and Discussion

I-V characterizations were conducted on the final SSG devices in a Lakeshore cryogenic probe station connected to a Keithley 4200SCS parameter analyzer and a Lakeshore 336 temperature controller. During the measurements, the chuck temperature was varied from 200 K to 300 K at intervals of 20 K. The device schematic with the associated wiring setup used during the measurements is given in Fig. 3a. Temperature dependent I-V measurements are presented in Fig. 3b, as current density versus voltage (J-V) plots. The graphs exhibit considerable temperature dependence in both positive and negative branches, indicating thermally assisted transfer of charge carriers. The J-V plots also show a clear asymmetry, i.e. a diode behavior with a high slope and a very small voltage at the lowest current levels. The observed rectification behavior can be attributed to the asymmetric electron barriers present at the Si/MoS$_2$ junction during the forward-biased and at the SLG/MoS$_2$ junction during the reverse-biased conditions. In addition, the current scales with device size (Fig. S4, in supporting information). A comparison of the room temperature I-V characteristics of the present devices with the results from "graphene/n-Si" Schottky diodes intensively studied by our research group[39–41] clearly indicates the electrical impact of MoS$_2$ in the present structures.

The measured forward biased I-V data was analyzed using the classic current transport models in semiconductor physics. In this regard, Fowler-Nordheim tunneling (FNT) [31], direct tunneling (DT) [42,43], space-charge-limited transport (SCL) [31], trap-assisted tunneling (TAT) [44] and thermionic emission (TE) [31] were examined for correlations with the experimental data. The strong temperature dependence that is evident from the measured J-V characteristics (Fig. 3b) rules out FNT and DT, as both conduction mechanisms should not exhibit temperature dependence. Moreover, the MoS$_2$ layer is thick enough (~15 nm) to suppress DT [43]. Beside FNT and DT, SCL is also ruled out due to the lack of a V$^2$ dependence in the current [45]. This was verified by analyzing the data using the FNT, DT and TAT models (not shown here) from which no correlations were found. This leaves TE as the most probable transport mechanism for current conduction across the SSG heterostructures. According to the conventional TE model (Eq.2), plots of "ln(J/T$^2$) versus 1/($k_B T$)" at given biases are expected to yield linear curves with negative slopes. The measured J-V data was replotted in this form and results in a set of Richardson's plots (Fig. 3c). The linearities observed in these plots strongly supports that current conduction across the SSG heterostructures results mainly from thermally stimulated transfer of charge carriers. The heterojunction emission-barrier height was extracted from the Richardson's plot and also from modeled I-V characteristics. The slope of each Richardson plot provides the respective activation energy, which in the current case corresponds to the heterojunction barrier height at 0 K, $\Phi_{Bo}(V)$, for a temperature independent α. To demonstrate how $\Phi_{Bo}(V)$ reacts to bias, the extracted activation energies are plotted as a function of voltage as in Fig. 3d, where a decrease in $\Phi_{Bo}$ is exhibited for increasing bias. This behaviour is a clear manifestation of a bias-dependent barrier lowering, which is an essencial part of the TE mechanism. As a demonstration of how the TE transport mechanism operates in the SSG structure, illustrative band diagrams are given in Figs. 4a and 4b in flat-band and forward-biased conditions, respectively.

The barrier heights were also obtained by modeling: The schematics in Figs. 5a and 5b demonstrate the charge carrier transfer in which thermally excited electrons leave the Si crystal and escape over the energy barrier at the Si/MoS$_2$ interface. The electron energy distributions in both materials are also illustrated in the schematics.

To calculate the I-V characteristics, it was assumed that the ballistic part of the electron transport from Si into the $MoS_2$ CB maximum is proportional to the total concentration of electrons in the energy range above the barrier maximum (i.e. hatched area in Fig. 5a). Ballistic transport is generally used for transport without scattering. In a more rigorous treatment, one would take scattering processes into account. Such attempts can be done in full detail only with more information about the interface than what is available for the present system. Based on the consistence of the data shown in the present work, we believe that the approximation used here is adequate. The energy distribution of electrons heading towards the barrier is given by the product of the Fermi-function and the density of states in the Si parabolic CB, which is proportional to $(E - E_F)^{1/2}$ for a Maxwellian electron gas. The I-V characteristics were then calculated for 200K (black), 260 K (blue) and 300 K (red), using a model (Eq. 1) analogous to the reasoning in [46], and fitted to the corresponding experimental data using bias-dependent barrier height functions as fitting parameters (Fig. 5c). The solid and dotted curves in this figure represent I-V characteristics calculated using Fermi- and Boltzmann-distributions, respectively. The Fermi-distribution fits the measured data better than the Boltzmann-distribution, especially in the higher bias range. As depicted in Fig. 5d, different "barrier height vs voltage" curves (black, blue and red) were needed to reproduce the measured I-V characteristics at 200 K, 260K and 300 K, respectively. A maximum difference of ~ 45 meV is noted at about 0.4 V between the shapes of the barrier curves which eventually merge for V ≥ ~1.2 V.

As described in the experimental section, the negative charge accommodated within the $MoS_2$ layer influences the shape of its CB and thereby limits the current across the SSG heterostructures. Assuming a Gaussian depth distribution of this charge as shown in Fig. 6a, the electric field, *F(x)*, was calculated for various voltage values using Eq. (6). The $MoS_2$ CB, $E_C$, was then calculated from the resulting *F(x)* values using Eq. (7). The black curves in Figs. 6b and 6c show the calculated $MoS_2$ CB demonstrating a band bending that increases with increasing bias. To account for the effect of image force lowering on the $MoS_2$ CB bending, calculations were done using Eq. 10 which includes Schottky lowering. The resulting CB bending curves are presented as the red curves in Figs. 6b and 6c. The maximum points of the CB curves were extracted from Fig. 6c and compared with the barrier function obtained by fitting the calculated I-V to the measured data at room temperature (RT). Fig. 6d presents a comparison in which the barrier data without image force lowering (black points) fits well, while the corresponding data with image force lowering (red points) do not.

The heterojunction barrier heights extracted by using Richardson's analysis from activation energies (Fig. 3c) and by calculating I-V curves and fitting them to the measured data (Fig. 5c) lead to different results: the values obtained from thermal activation data (Fig. 3d) are considerably smaller than those from I-V fitting (Fig. 5d). This aspect requires further discussion: At barrier heights larger than ~$3kT$, the Fermi-distribution can be approximated by the Boltzmann-distribution to obtain the Richardson's expression (Eq. 2). However, as shown in Fig. 5a, the Boltzmann-distribution assumes a larger electron concentration than actually is available at very low energy barriers and thus overestimates the current. Therefore, this approximation would be in jeopardy for barriers lower than ~$3kT$, which corresponds to the red dashed line in Figs. 5a and 5b. As depicted in Fig. 5c, the calculated current based on Boltzmann-distribution diverges from the measured data at high voltages, while the data established by Fermi distribution show a better agreement. The discrepancy between the current calculated using the Boltzmann-distribution and the measured data for V < 0.4 V might originate from the assumption on the electron transition probability and density of states as pointed out in the experimental section.

Approximating the Gaussian distribution of the MoS$_2$ charge (Fig. 6a) by a sheet charge and multiplying its full-width-half-maximum by its amplitude, one would obtain a concentration of about 2.8x10$^{16}$ m$^{-2}$. Moving this charge between the Si/graphene and Si/MoS$_2$ interfaces would introduce a voltage variation of ~1.4 V across the MoS$_2$ layer. This is in the range of the flat band voltage ($V_{FB}$) shift due to negative charges in the MoS$_2$ bulk observed in our earlier work [26,47]. It is worth noting here that, for a non-homogeneous charge distribution across the SSG structure (Fig. 2c), the electric field at the Si/MoS$_2$ interface differs considerably from $V/d$. As demonstrated in Figs. 6b and 6c, the negative charges in MoS$_2$ induce a barrier (EB2) whose maximum point gradually shifts towards the Si/MoS$_2$ interface for increasing applied voltage. The height of this barrier also decreases with increasing voltage and eventually disappears as can be noticed from the figures. As electrons leave Si and approach the Si/MoS$_2$ interface, they may experience an attraction force from the image charges that build up in the Si. An image force associated with these charges can slightly lower the effective height of the heterojunction barrier (EB2) and the lowering may increase with increasing bias as demonstrated by the red curves in Figs. 6b and 6c. However, the comparison in Fig. 6d asserts that the charge induced barrier (EB2) dominates the interface barrier (EB1) and determines the current conduction across the SSG structure. Also, this result ratifies that image force lowering may not be applicable to the present case [48].

From the differences exhibited among the barrier height values obtained by modelling the I-V data at different temperatures (Figs. 7a and 7b), we found that the heterojunction barrier height, $\Phi_B$, is temperature dependent. As shown in Figs. 7c and 7d, $\Phi_B$ has a nearly linear temperature dependence that can be formulated using a first order approximation as in Eq. 3. The proportionality factor $\alpha$ in Eq. 3 could be related to the change in entropy taking place in the electron ensemble while transferring from Si to MoS$_2$. According to Eqs. 3 and 4, barrier height values obtained from activation plots based on the conventional Richardson's expression can be identified as $\Phi_{B0}$, which corresponds to the intercepts of the "$\Phi_B(T)$ vs $T$" plots shown in Figs. 7c and 7d. Other possibilities to explain the temperature dependence of the barrier height could be one or more of the following: change in the difference between the band gaps of Si and MoS$_2$, strain at the Si/MoS$_2$ interface and/or barrier inhomogeneity across the sample area. As reported in previous works, strain influences the effective mass of electrons in MoS$_2$ [49], which makes its contribution to the observed temperature dependence under discussion a possibility.

Throughout the discussions in the paper, the overall charge carrier transfer between Si and MoS$_2$ has been considered unaffected by the interfacial silicon oxide layer lying between the two materials as revealed by the TEM cross-section (Supplementray information, Fig. S3). This layer is found to be very leaky compared to standard SiO$_2$ of similar thickness and bias level [50], leading to the assumption that it is nearly transparent to electrons under DC bias. This assures that the bottle neck for the current conduction across the SSG structure is therefore the charge carrier injection across the MoS$_2$ layer.

IV. **Conclusions**

Semiconductor-semiconductor-graphene (SSG) vertical heterostructures were investigated with respect to the vertical current conduction mechanism in order to assess their potential as the emitter diodes in vertical hot electron transistors. The SSG structures were fabricated using a scalable and CMOS compatible process. The measured electrical data exhibits asymmetric I-V characteristics with clear temperature dependence. Richardson's plots of the measured data confirmed thermionic emission as the main conduction mechanism, which is the desired mechanism for high performance hot electron transistors. This result was confirmed through extensive analytical modelling of the I-V characteristics. The zero-bias heterojunction barrier height was extracted by both methods. The value obtained through the former method is considerably smaller than that from the latter. In fact, the heterojunction barrier height was found to have a linear dependence on temperature, and the values obtained from Richardson's plots appear to correspond to values at T = 0 K. The presented extraction method and model may serve as a guideline for future experiments on electronic properties of 2D heterostructures. In particular, the presented data suggests $MoS_2$ as a thermionic emission barrier material for vertical hot electron transistors.

**Acknowledgements**

The authors thank Gregor Schulte from University of Siegen for his kind help in the deposition of initial Molybdenum films. The authors are also thankful to Prof. Joachim Mayer and Maximilian Kruth for TEM imaging. Financial support from the European Commission (Graphene Flagship, 785219; QUEFORMAL, 829035) and the German Ministry of Education and Research, BMBF (GIMMIK, 03XP0210; NEUROTEC, 16ES1134) is gratefully acknowledged.

**Supporting Information:**

This paper is accompanied by a "supporting information" document containing the following contents:

- **Fig. S1**: Illustrative schematic diagrams showing the structure and operation principle of a graphene-based hot electron transistor (GBT) in which the SSG device demonstrated in this paper is used as the active emitter diode (Page S-2);

- **Fig. S2**: AFM investigation on the as-grown $MoS_2$ film showing the actual thickness of the $MoS_2$ film after grown (Page S-3);

- **Fig. S3:** TEM investigation of $MoS_2$ revealing the polycrystalline nature of the film (Page S-4);

- **Fig. S4:** A brief discussion on the area scaling of the measured I-V characteristics and the corresponding result (Page S-5).

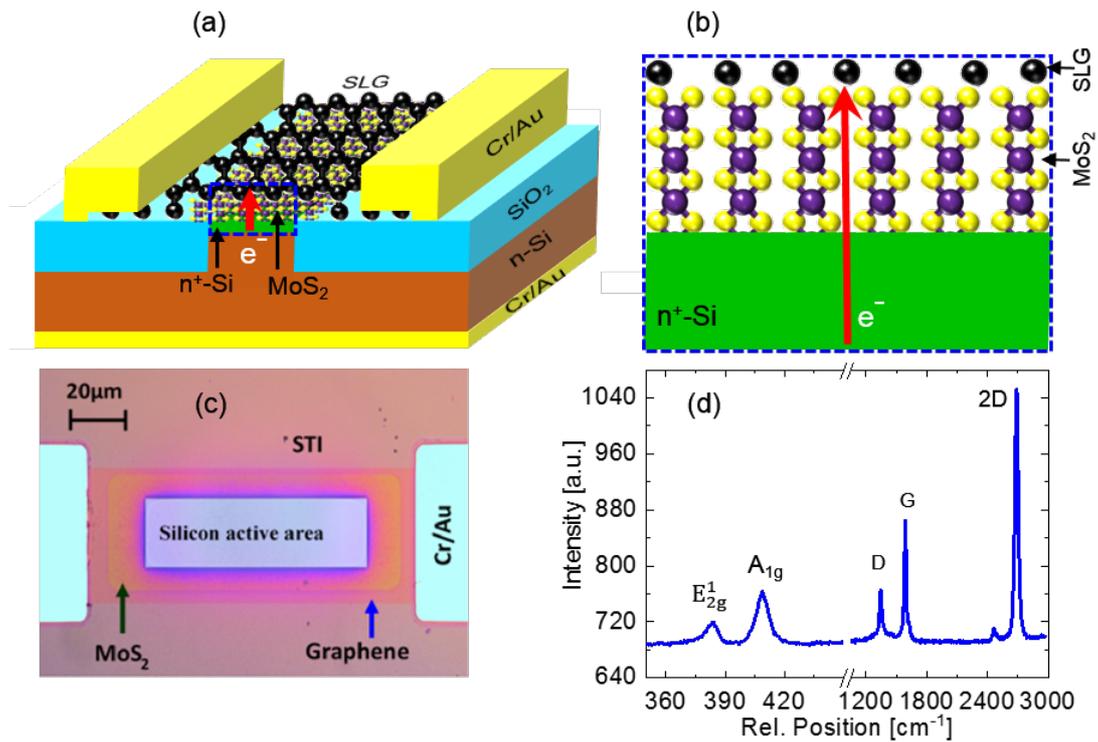

Figure 1. (a) Schematic of an isometric view of the as-fabricated SSG structure, where MoS$_2$ is used as an emission barrier and the red vertical arrow indicating the electron transport direction. (b) Magnified view of the SSG heterojunction in the active region marked by the blue dashed box in "a". (c) Optical micrograph of top-view of the actual SSG device, and (d) Raman spectrum of the MoS$_2$ and single layer graphene. The E$^1_{2g}$ and A$^1_g$ peaks indicate the MoS$_2$ phase formation, whereas an intense 2D band with 2D/G intensity ration > 1 confirms the presence of a single layer graphene of reasonable quality.

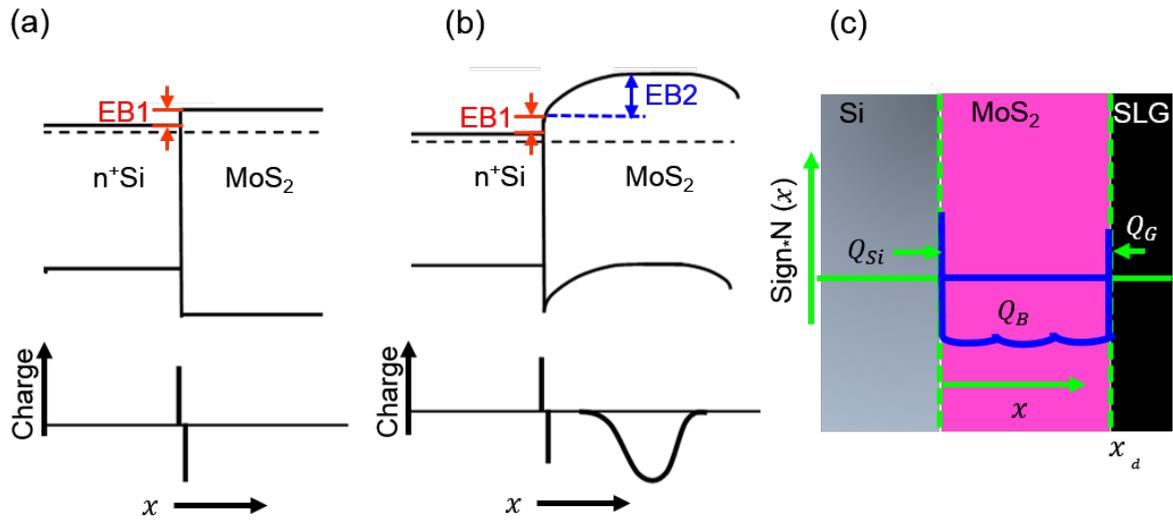

Figure 2. Schematic diagram illustrating the geometrical representation of the Si-MoS$_2$ band alignment and the associated charge distribution for (a) a neutral MoS$_2$ layer and (b) MoS$_2$ with negative charge in its bulk. (c) Schematic diagram for a simple demonstration of the charge distribution in the SSG structure.

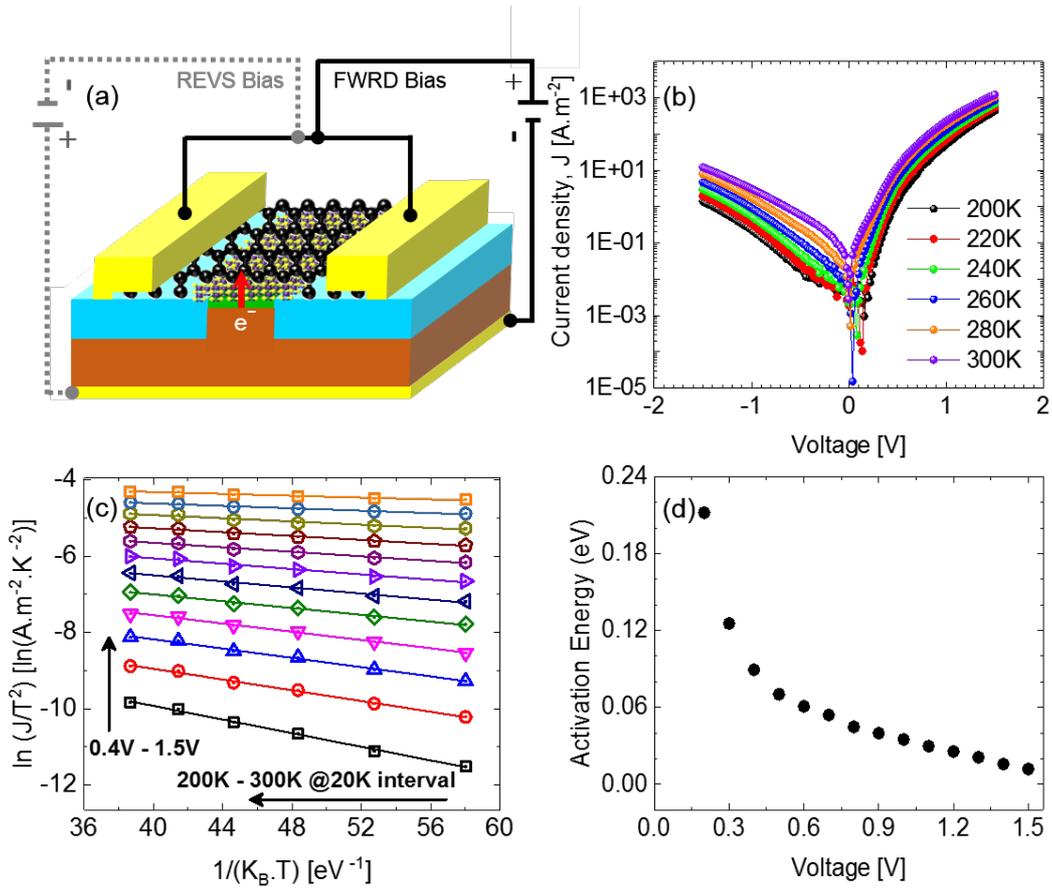

Figure 3. (a) Schematic of isometric view of the SSG structure with a wiring setup used for the I-V measurements. (b) I-V characteristics measured on the SSG devices in vacuum at temperatures ranging from 200 K to 300 K with a 20 K interval showing asymmetric and temperature dependent current with low turn on voltages ($V_{ON}$). (c) $\ln[J/T^2]$ vs $(K_B.T)^{-1}$, Richardson's plots with high linearities indicating a major contribution of thermionic emission to the overall current across the SSG devices. (d) Activation energy, $E_a$, vs $\sqrt{V}$ plot from which the zero-bias heterojunction barrier height, $\Phi_{BO}$ value is extracted.

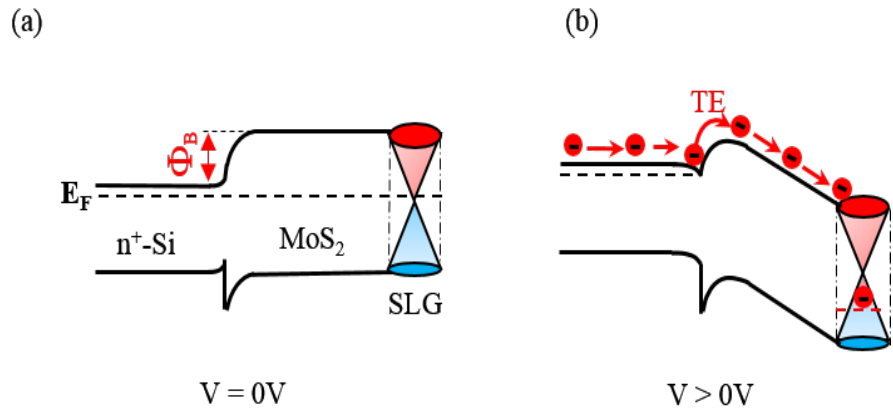

Figure 4. Schematics of band diagrams of an SSG structure in: a) thermal equilibrium and b) forward biased conditions with illustrations of the electron transfer process from Si to MoS$_2$ via thermionic emission (TE).

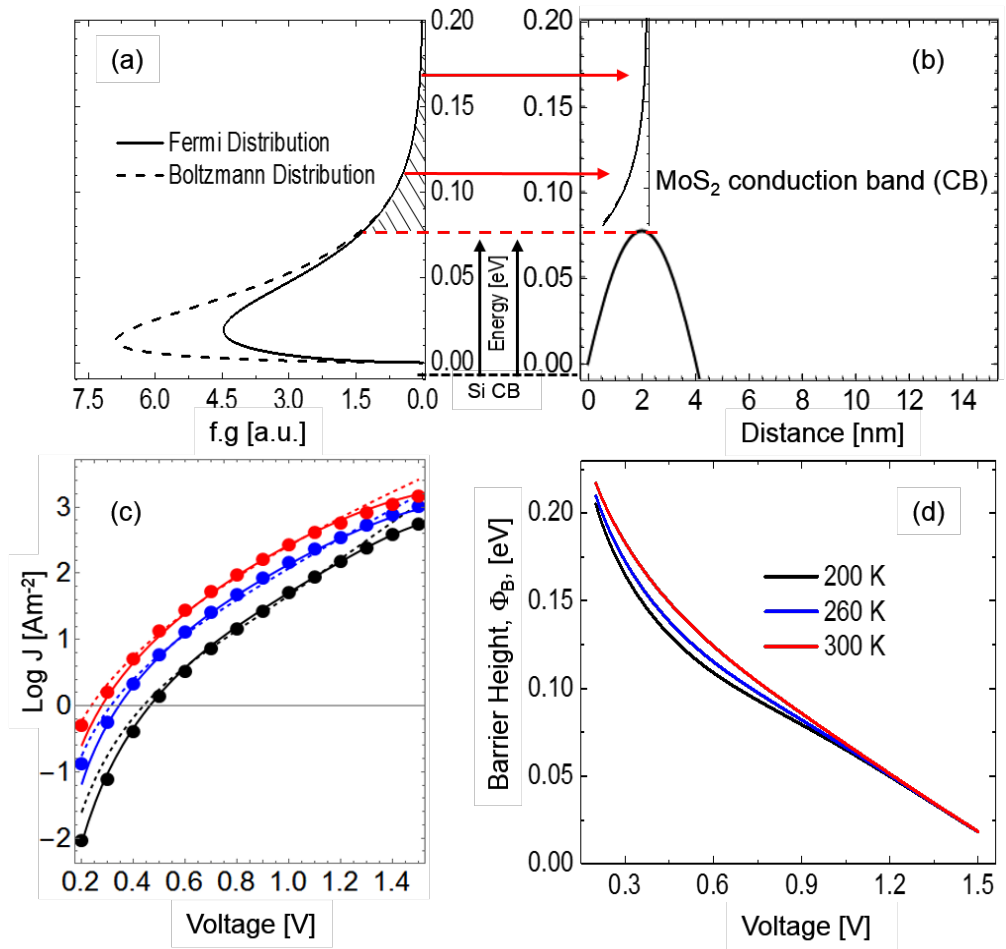

Figure 5. Boltzmann versus Fermi-Dirac statistics for low energy barriers: (a) "f.g vs E" graph showing distribution of carriers at the Si CB edge. (b) Graph illustrating the $MoS_2$ CB bending as a function of distance into the $MoS_2$ depth. The schematic on top of the highest potential (EB2) is the Fermi tail of distribution of electron states in the $MoS_2$, resembling that of Si at similar energy values. The horizontal red arrows illustrate the transfer of electrons from Si to $MoS_2$, while the dashed line marks the barrier maximum to to indicate that only electrons with energies above this line will be injected into $MoS_2$. The hatched region shows the concentration of electrons that can be injected into $MoS_2$. (c) Comparison of measured I-V (dots) and simulated I-V (lines) for 200 K (black), 260 K (blue) and 300 K (red) measurement temperatures. The solid I-V curves in (c) are calculated using Fermi-distribution, while the dashed ones are based on Boltzmann-distribution. (d) Comparison of barrier functions used to calculate I-V curves and fit them to the measured data at different temperatures as shown in (c). The barrier curves are seen to merge for V ≥ 1V.

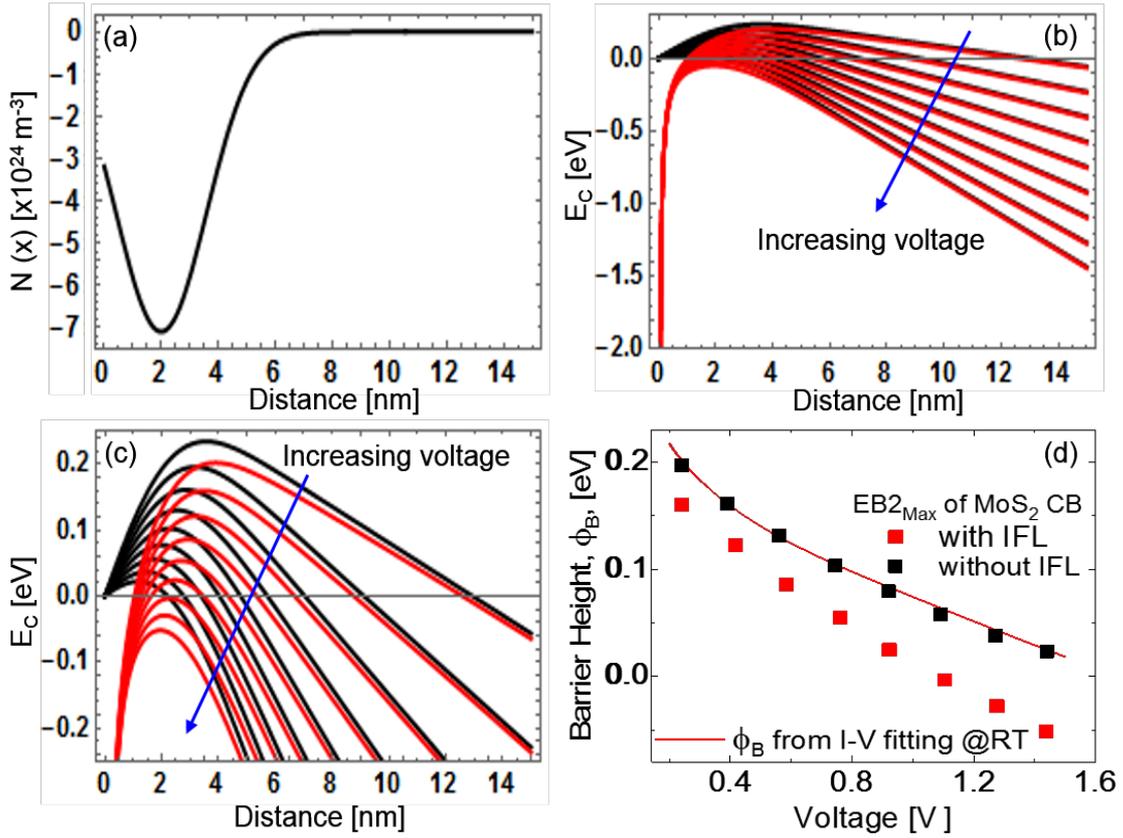

Figure 6. (a) Gaussian distribution of negative charges inside the MoS$_2$ bulk. (b) MoS$_2$ CB bending as a function of voltage across the layer depicting the formation of the electron barrier EB2 which is determined by the negative charge in the bulk. (c) Magnified version of the graph in (b) along the vertical axis for a better visibility of the CB maxima in determining the value of EB2 Max at different voltages. (d) Comparison of the maxima of the potential barrier plots in (c) (symbols) and the barrier function used to fit the calculated I-V with the measured data at 300 K (red solid curve) entailing that the charge induced barrier (EB2) determines the current across the present structures. The red and black curves and points represent calculations with and without consideration, respectively, of the effect of image force barrier lowering (IFL). The parameters used to calculation the MoS$_2$ CB are: $x_d$ = 15, $Q_B$ (concentration of charge in the MoS$_2$ bulk) = -7.097x10^24 m$^{-3}$, $\varepsilon = 3$, and $\Phi_G - \Phi_{Si} = \sim 0.45\ eV$.

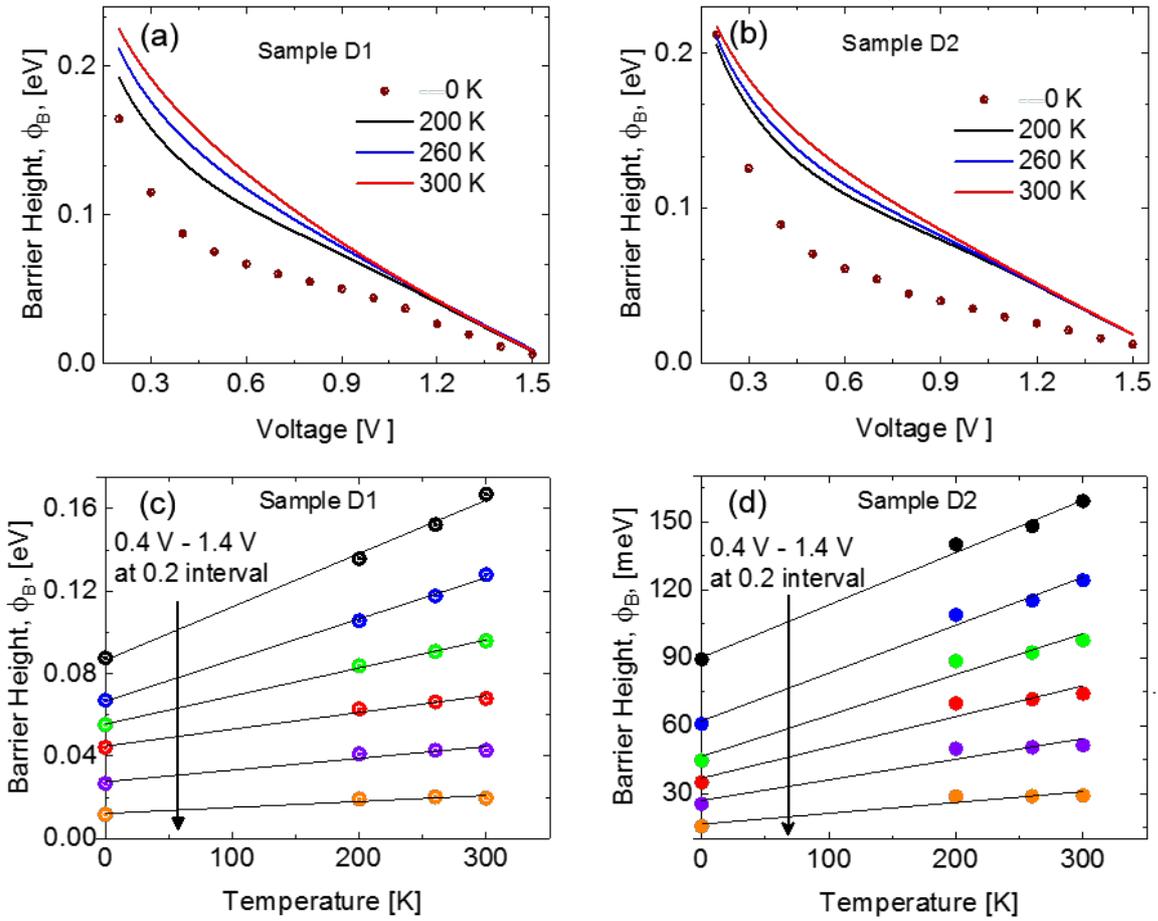

Figure 7. Comparison of barrier height values extracted from activation plots (symbols) with barrer height functions used to fit the measured I-V curves at 200K (black), 260K (blue) and 300K (red) for (a) sample D1 and (b) sample D2. Barrier height, $\Phi_B$, as a function of temperature, T, at different voltages for (c) sample D1 and (d) sample D2 showing a linear dependence of $\Phi_B$ on temperature.